# Analysing software failure using runtime verification and LTL

Zahra Yazdanparast

*School of Electrical and Computer Engineering,*
*Tarbiat Modares University,*
*Tehran, Iran*
zahra.yazdanparast@modares.ac.ir

## Abstract

A self-healing software system is an advanced computer program or system designed to detect, diagnose, and automatically recover from faults or errors without human intervention. These systems are typically employed in mission-critical applications where downtime can have significant financial or operational consequences. Failure detection is one of the important steps in the self-healing system. In this research, a method using runtime verification is proposed to diagnose four types of errors at the component level. The simulation on mRUBIS shows that the suggested method has the necessary efficiency in detecting the occurrence of failures.

## 1. Introduction

A self-adaptation system is able to modify its behavior or structure in response to changes in its operational environment or unwanted changes created in the system [1]. In other words, a self-adaptation system continuously evaluates its behavior during execution, and if these behaviors violate the goals of the system, the system changes its behavior to achieve the goals again [2]. Self-healing is a self-adaptation feature in which the system automatically detects and repairs hardware and software problems. The main goal of self-healing is to have an automatic system that can heal itself without human intervention. This system has predefined actions and procedures that are suitable for recovering the system from different failure modes. Such a system can change the state of the system from failed to healthy [3].

The terms self-healing and self-repair are often used interchangeably, but they are different. Self-healing at the time of execution and converting an unsuccessful execution into a successful execution is done at the service and component level. However, self-repair is in the design and testing stage, and changes are made at the source code level to fix the error and prevent it from happening again [4, 5].

Runtime models, a promising way to manage complexity in runtime environments, are the development of adaptive mechanisms that use software models. Work on runtime models seeks to expand the application of models generated in model-driven engineering (MDE) methods in the runtime environment.

By using runtime models, design errors can be fixed or new design decisions can be made in the running system to support running controlled design. Runtime models make an important contribution to the field of automated computing and provide meta-information to guide autonomous decision-making. MAPE architecture is one of the most important models used in self-healing, which consists of four components: monitoring, analysis, planning, and execution. In this research, a method for failure analysis using runtime verification is proposed.

## 2. Related work

Software architecture is important in self-healing systems to represent the system structure and decide on the levels of healing. One of the things that the self-healing system should always be aware of is its architecture, and in general, software architecture should always be considered to provide a solution in the design of self-healing systems.

In 2019, [6] presented a microservice architecture to solve the Docker congestion cluster problem in the cloud. The proposed method suggests self-adaptation by following the MAPE-K model. The main contribution of this method is to use the utility function in the adaptation process.

Vogel et al. [7] has presented the mRUBiS method based on a self-adaptation model-based architecture. In this article, the MAPE architecture is used for the adaptive engine part. To express the execution time model of adaptive software architecture, a general and simple modeling language called CompArch (Component Architecture) has been developed. This method uses the utility function to evaluate self-adaptation. mRUBiS is useful as it supports early testing and evaluation of solutions.

The Rainbow Architecture [8-10] is a self-healing framework that uses a top-down focused style. This architecture is divided into an architecture layer and a system layer (including managed resources). The architecture layer has various components that are responsible for defining self-healing programs. A manager provides a set of constraints and recovery plans that the service uses to evaluate system behavior. The evaluations have been performed using a three-tiered abstract architectural model that autonomously classifies system behaviors.

The MAPE-K architecture [11, 12] is used for the design and implementation of the adaptation manager part in adaptation systems, which was introduced by IBM for automated computing systems. This control loop continuously monitors the system and its operating environment. Then, the information obtained from the monitoring is analyzed and if changes are needed, the necessary plans are made and implemented on the system. All four parts of monitoring, analysis, planning, and execution have access to common knowledge to improve their performance. The MAPE-K control loop is used to design autonomous elements. Autonomous systems are created from the interaction of a set of autonomous elements, and the management of their internal behavior as well as the management of interactions between these elements is done based on specific policies.

## 3. Proposed method

In the monitor section, the required information on events is collected and sent to the analysis where events are analyzed using linear time logic (LTL) to determine the type of failure. By using the information received from monitoring, the occurrence of the failure is detected. The types of considered failures include the following:

- The component is in unknown status (CF1).
- The number of component exception failures is greater than the specified threshold (CF2).
- Component is removed from the architecture (CF3).
- The connection between two components is broken (CF4).

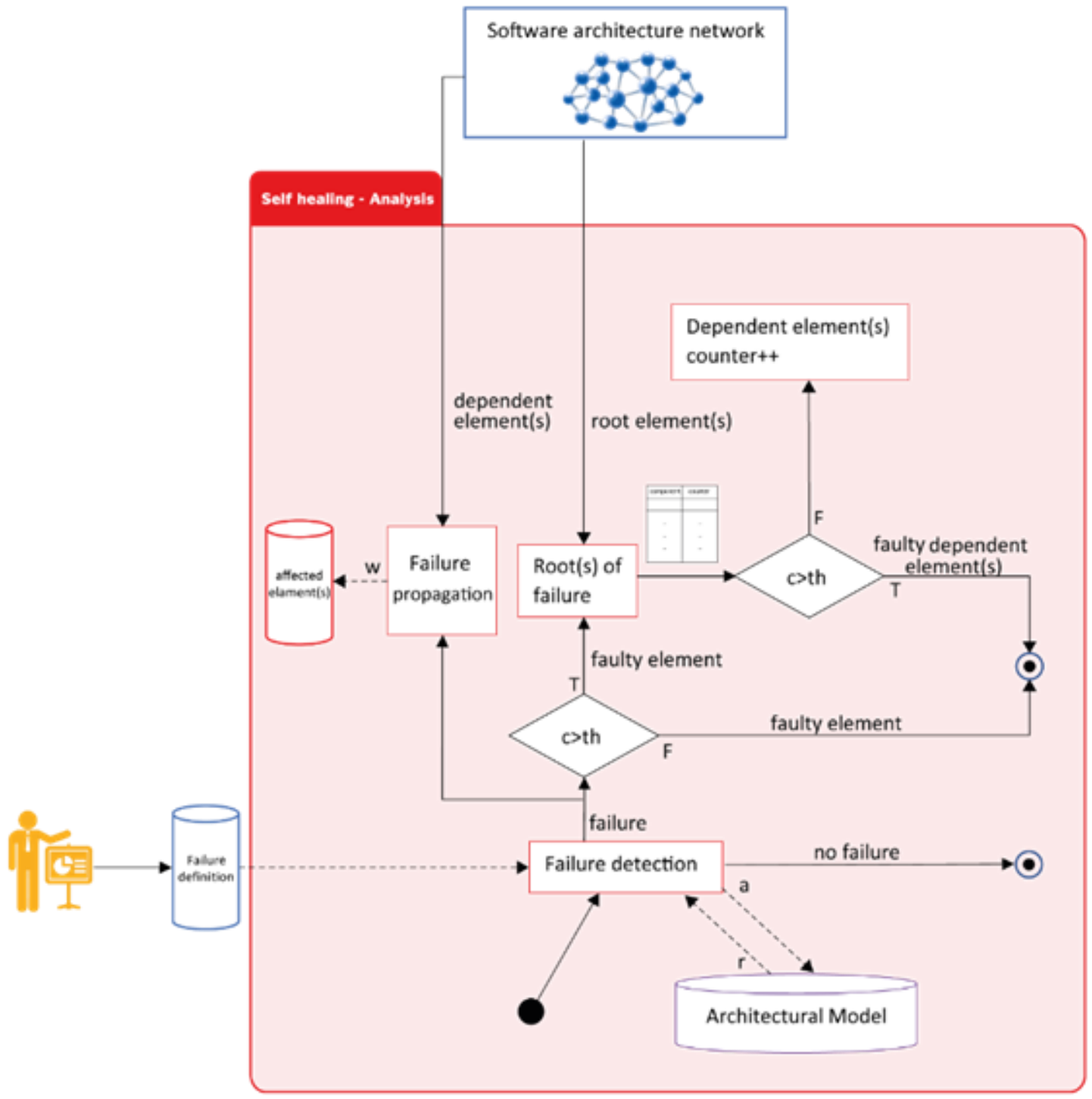

Figure 1: Analysing failure.

The occurrence of each failure is detected using a runtime verification approach. First, write an LTL formula for each failure, and the FSM corresponding to that formula is automatically generated. Then, using the expected code of each FSM, we detect the occurrence of each failure and its type at runtime.

There is also a counter per component that counts the number of times a component fails. If the number of component failures exceeds the threshold, further analysis is required. Because the root of the failure may be the components dependent on this component, and the cause of the failure is not the component itself.

### 3.1. Expression of types of failures in LTL language

In this section, the types of failures using LTL are described, which are introduced in Table 1.

Table 1: Expression of failures in LTL.

| **Expression of crashes in LTL language** | **Crash number** |
|---|---|
| $\varphi_1 = G\ (isUnknown)$ | 1 |
| $\varphi_2 = G$ (isStarted && lowException) | 2 |
| $\varphi_3 = G$ (isStartedComponent1 && isStartedComponent2 && connector) | 3 |

### 3.2. Conversion of types of failures from LTL to FSM

At this stage, the types of failures expressed in LTL language are converted to FSM using ltl3tools. In the first type of failure, the component should not be in the unknown state. It can be seen in FSMCF1 that if the component is in an unknown state, the system state will enter the broken state.

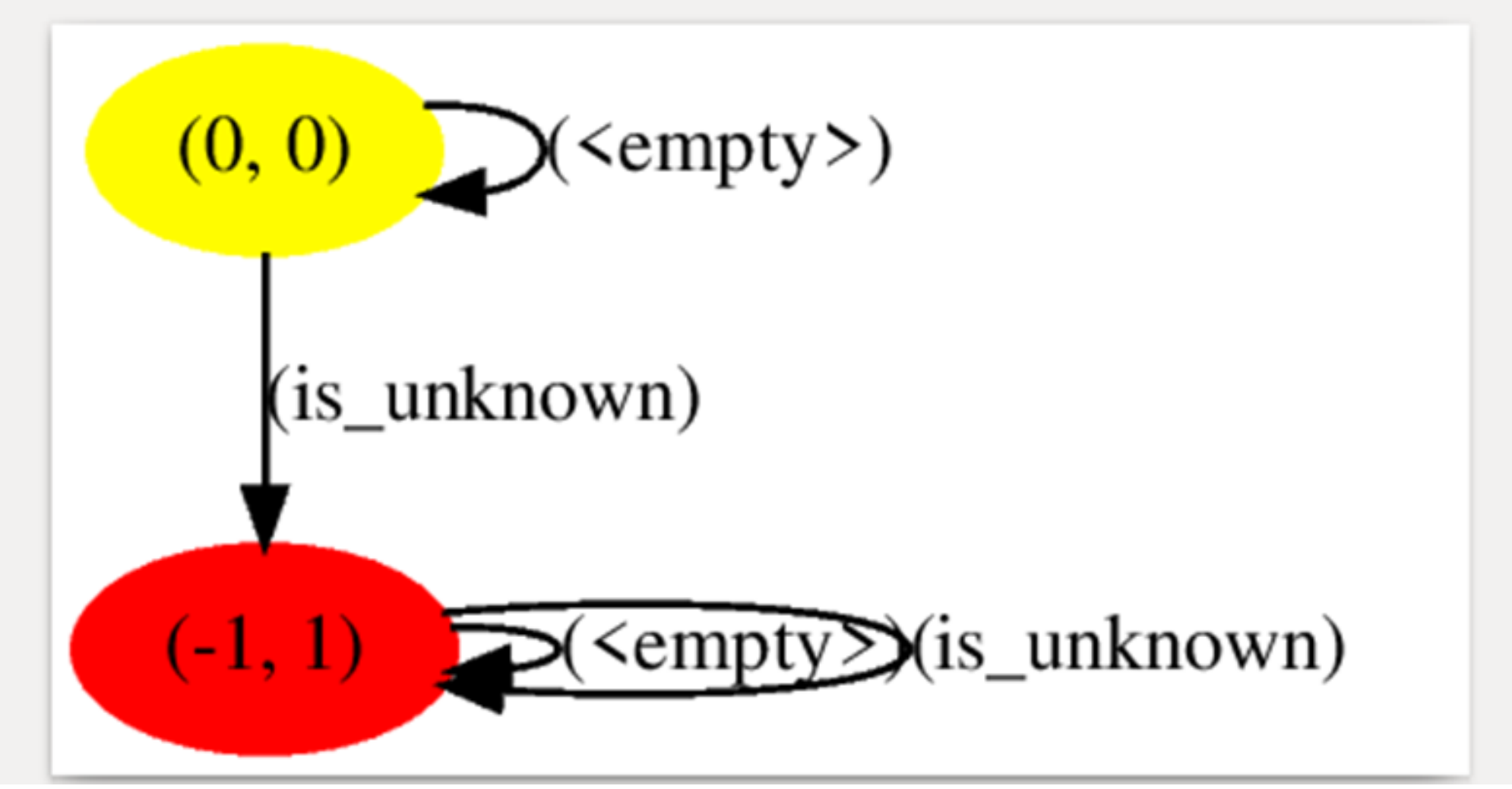

Figure 2: FSM of the first failure.

The second failure checks the number of exception failures per component. Exception failures should not exceed a predefined threshold.

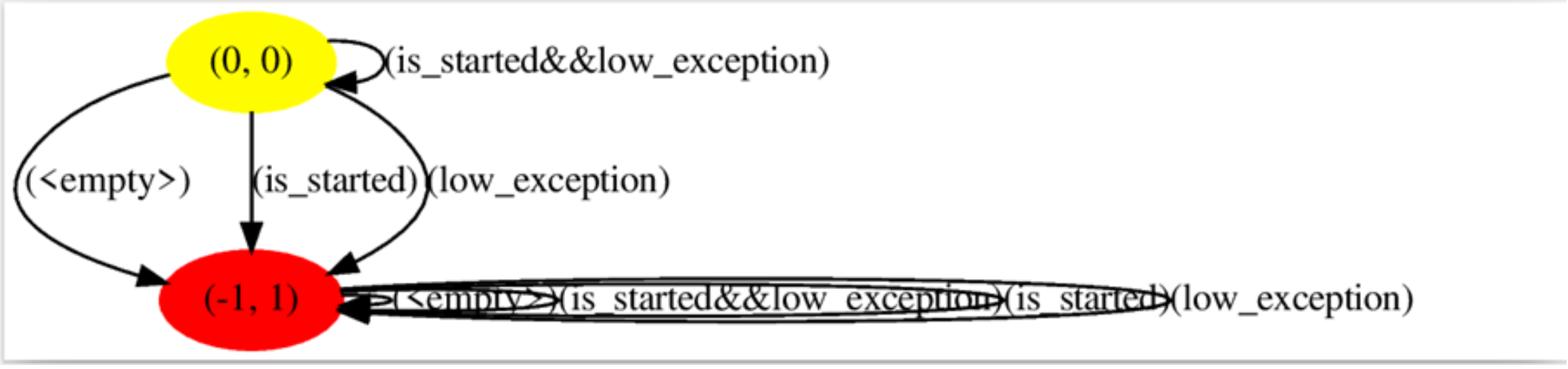

Figure 3: FSM of the second crash.

The third failure checks whether the connection between the two components is established.

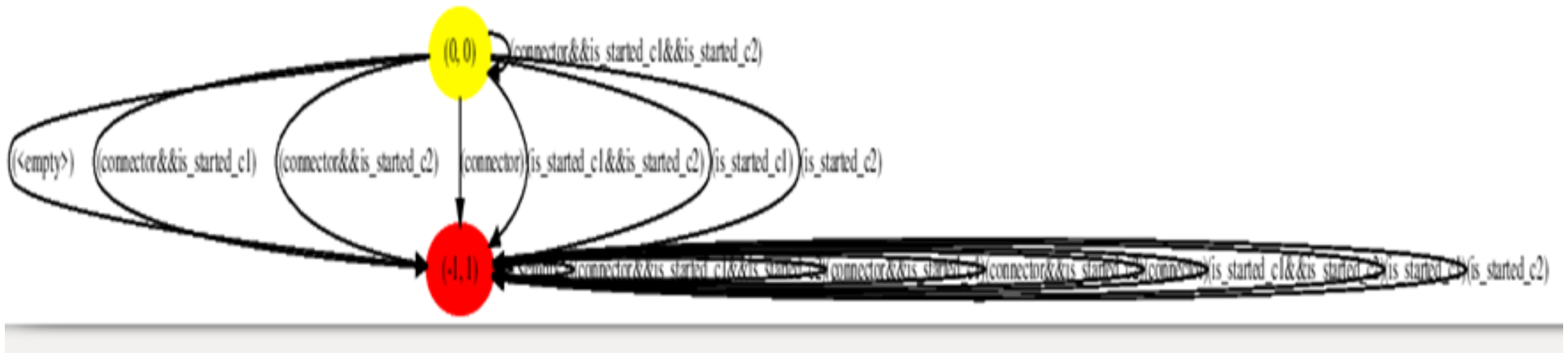

Figure 4: FSM of the third crash

## 4. Evaluation of results

In order to evaluate the proposed method, it is necessary to inject failures into the simulator. The mRUBiS simulator has failure injection capability and reflects failures in the architectural runtime model.

In the implementation scenario, failures are randomly selected among the existing failures and injected into the architectural model. The first failure is injected into the architecture. The self-healing loop is executed. First, in the analysis, the number of failures of this component is checked. If the number of times the component is damaged is less than the threshold, the same component is declared damaged and the type of crash is determined. If the component has dependencies and the number of failures of the component is more than the threshold, it is necessary to check the dependent components. Because the root of the failure may be dependent components. Dependent components that have been declared as broken dependencies more than a threshold are identified. If the failure of several dependent components exceeds the threshold, the most important dependent component is selected as the final dependent component based on its utility function and declared as a failed component. Since the threshold of an adaptive form is determined, one is subtracted from the threshold of the failed component, so in case of a fail again, it will be dealt with sooner.

### 4.1. Investigating dependent components

Suppose component x is declared failed. Component x uses the output of other components as its input. Therefore, the root of the failure may be one of the dependent components and the x component itself is not damaged. In the first scenario, in a simple way and using the graph of components, only dependent components are identified. However, the same component x that is declared failed will be repaired.

The first type of failure is applied to the Item Management Service component. The failure of this component is detected. Then, by using the graph of the components, the dependent components are identified. Item Management Service component uses the output of three components Authentication Service, Persistence Service, and Query Service. Therefore, there is a possibility that the cause of the failure is at least one of these three components. A variable named DependentFailureCount is defined whose value is incremented each time a dependent component is declared. For example, here, the Authentication Service component was identified as a dependent component for the broken Item Management Service component. So one is added to its counter. In this scenario, only dependents are detected. Nevertheless, the Item Management Service component is considered a failed component and is healed.

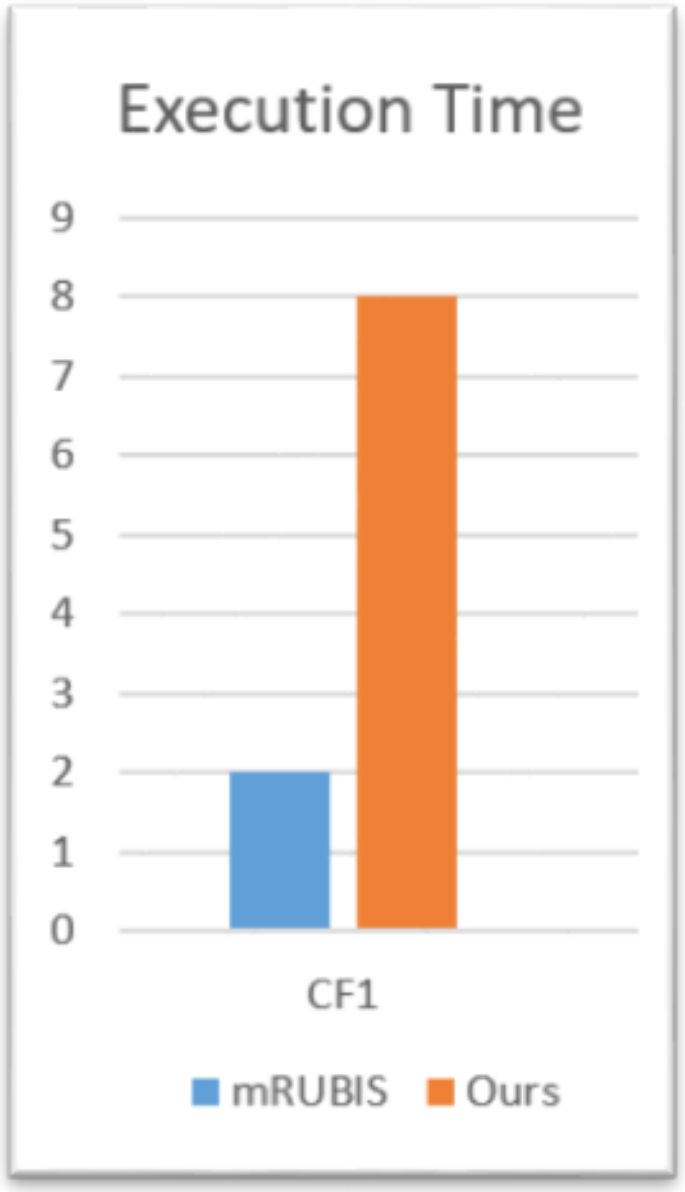

Figure 5: The result of the first scenario.

### 4.2. Use of runtime verification and LTL to detect failures

In this scenario, runtime verification is used to detect the failure. Other parts of MAPE are unchanged and only LTL has been added to the analysis part. LTL is implemented using aspectj and added to the mRUBIS emulator. As shown in Figure 6, four failures are injected into the simulator. Each of the graphs shows the execution time of injecting a failure type and executing the self-healing loop.

For example, CF2 failure occurs for the Query Service component. During this failure, the interface of this component is disconnected. The occurrence of this failure is detected using runtime verification and aspectj modes. Then, the component is restarted to heal.

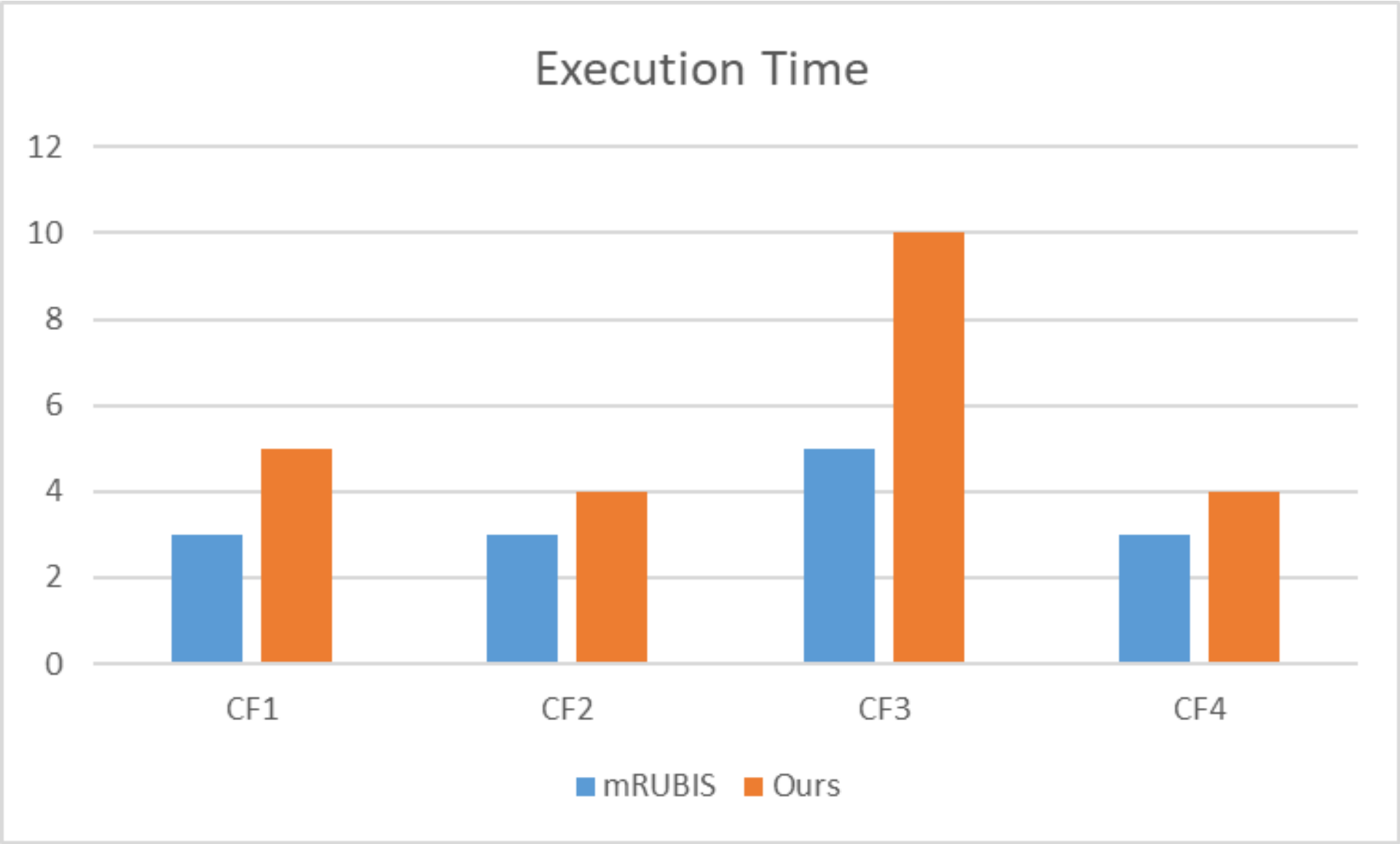

Figure 6: The result of the second scenario.

## 5. Conclusion

In this study, the analysis of software failures was investigated using runtime verification. Runtime verification examines an implementation of the system and determines the correctness or violation of a feature. Four types of failures were considered and written in LTL language and their FSM was drawn to detect the occurrence of failures. Automating the identification and correction of software glitches, self-healing systems offer substantial reductions in downtime, bolster system dependability, and boost operational effectiveness. Nevertheless, creating and deploying these systems necessitates thoughtful evaluation of elements like system intricacy, scalability, and possible unforeseen outcomes.

## References

[1] R. De Lemos *et al.*, "Software engineering for self-adaptive systems: A second research roadmap," in *Software Engineering for Self-Adaptive Systems II: International Seminar, Dagstuhl Castle, Germany, October 24-29, 2010 Revised Selected and Invited Papers*, 2013: Springer, pp. 1-32.

[2] M. Salehie and L. Tahvildari, "Self-adaptive software: Landscape and research challenges," *ACM transactions on autonomous and adaptive systems (TAAS)*, vol. 4, no. 2, pp. 1-42, 2009.

[3] A. A. Hudaib, H. N. Fakhouri, F. E. Al Adwan, and S. N. Fakhouri, "A survey about self-healing systems (desktop and web application)," *Communications and Network*, vol. 9, no. 01, pp. 71-88, 2017.

[4] R. Frei, R. McWilliam, B. Derrick, A. Purvis, A. Tiwari, and G. Di Marzo Serugendo, "Self-healing and self-repairing technologies," *The International Journal of Advanced Manufacturing Technology*, vol. 69, pp. 1033-1061, 2013.

[5] L. Gazzola, D. Micucci, and L. Mariani, "Automatic software repair: A survey," in *Proceedings of the 40th International Conference on Software Engineering*, 2018, pp. 1219-1219.
[6] B. Magableh and M. Almiani, "A self healing microservices architecture: A case study in docker swarm cluster," in *Advanced Information Networking and Applications: Proceedings of the 33rd International Conference on Advanced Information Networking and Applications (AINA-2019) 33*, 2020: Springer, pp. 846-858.
[7] T. Vogel, "mRUBiS: An exemplar for model-based architectural self-healing and self-optimization," in *Proceedings of the 13th International Conference on Software Engineering for Adaptive and Self-Managing Systems*, 2018, pp. 101-107.
[8] D. Garlan, "Invited Talk: Rainbow: Engineering Support for Self-Healing Systems," in *2009 XXIII Brazilian Symposium on Software Engineering*, 2009: IEEE, pp. xiv-xiv.
[9] D. Garlan, S.-W. Cheng, A.-C. Huang, B. Schmerl, and P. Steenkiste, "Rainbow: Architecture-based self-adaptation with reusable infrastructure," *Computer,* vol. 37, no. 10, pp. 46-54, 2004.
[10] S.-W. Cheng, D. Garlan, and B. Schmerl, "Architecture-based self-adaptation in the presence of multiple objectives," in *Proceedings of the 2006 international workshop on Self-adaptation and self-managing systems*, 2006, pp. 2-8.
[11] J. O. Kephart and D. M. Chess, "The vision of autonomic computing," *Computer,* vol. 36, no. 1, pp. 41-50, 2003.
[12] A. G. Ganek and T. A. Corbi, "The dawning of the autonomic computing era," *IBM systems Journal,* vol. 42, no. 1, pp. 5-18, 2003.